\documentclass[journal]{IEEEtran}

\usepackage{amsmath}
\usepackage{amssymb} 
\usepackage{cases}

\ifCLASSINFOpdf
   \usepackage[pdftex]{graphicx}
   \graphicspath{{./img/}}
   \DeclareGraphicsExtensions{.pdf,.jpeg}
\else
\fi
\usepackage{amsmath}
\begin{document}
\title{Implementing High-Order FIR Filters in FPGAs}
\author{Philipp~F\"odisch,
        Artsiom~Bryksa,
				Bert~Lange,
				Wolfgang~Enghardt
        and~Peter~Kaever

\thanks{P.~F\"odisch, B.~Lange and P.~Kaever are with the Department of Research Technology, Helmholtz-Zentrum Dresden - Rossendorf,  Germany (e-mail: p.foedisch@hzdr.de).}
\thanks{A.~Bryksa is with the Faculty of Radiophysics and Computer Technologies, Belarusian State University.}
\thanks{W.~Enghardt is with OncoRay - National Center for Radiation Research in Oncology, Faculty of Medicine and University Hospital Carl Gustav Carus, Technische Universit\"at Dresden; Institute of Radiooncology, Helmholtz-Zentrum Dresden - Rossendorf, Germany and German Cancer Consortium (DKTK) and German Cancer Research Center (DKFZ), Germany.}}


\maketitle

\begin{abstract}
Contemporary field-programmable gate arrays (FPGAs) are predestined for the application of finite impulse response (FIR) filters. Their embedded digital signal processing~(DSP) blocks for multiply-accumulate operations enable efficient fixed-point computations, in cases where the filter structure is accurately mapped to the dedicated hardware architecture. This brief presents a generic systolic structure for high-order FIR filters, efficiently exploiting the hardware resources of an FPGA in terms of routability and timing. Although this seems to be an easily implementable task, the synthesizing tools require an adaptation of the straightforward digital filter implementation for an optimal mapping. Using the example of a symmetric FIR filter with 90 taps, we demonstrate the performance of the proposed structure with FPGAs from Xilinx and Altera. The implementation utilizes less than 1\,\% of slice logic and runs at clock frequencies up to 526\,MHz. Moreover, an enhancement of the structure ultimately provides an extended dynamic range for the quantized coefficients without the costs of additional slice logic.
\end{abstract}
\begin{IEEEkeywords}
digital filters, field-programmable gate arrays, FIR filters, fixed-point arithmetic
\end{IEEEkeywords}

\IEEEpeerreviewmaketitle

\section{Introduction}
\IEEEPARstart{N}{owadays}, finite impulse response (FIR) filters are a major application of field-programmable gate arrays (FPGAs) in the context of digital signal processing (DSP). An FPGA design is traditionally implemented in a hardware description language on the register-transfer level (RTL). Therefore, the design flow from the RTL top level down to the logic and circuit level postulates clear definitions and constraints for an optimal implementation result. As a consequence, the architecture of the device impacts the design, and a straight separation of the abstraction layers is practically impossible for an efficient realization in terms of logic utilization or speed. Although some approaches to hardware efficient filter structures exist \cite{meher,park}, the key to a valuable exploitation of the hardware resources is an adaptation of the FIR filter structure to the architecture of the FPGA. Moreover, at the early stage of filter design, a reduction of hardware complexity is possible \cite{mehrnia2}. At last, an adaptation of the mathematical modelling to the dedicated silicon (DSP blocks) maximizes performance \cite{xilinx_dsp} while versatility decreases.

In fact, an implementation of numerical algorithms in an FPGA, including FIR filters, is a trade-off between targeted precision, allocated logic, achievable clock frequency, and allowed latency. An operation with a fixed-point arithmetic is usually preferred if resources or speed weigh more than the highest precision.
Thus, state-of-the-art FPGAs have hundreds of built-in DSP blocks, capable of fixed-point operations with a precision of at least $18\,\mathrm{bits}$ \cite{altera_wp,dsp48e}. Consequently, we will adapt a symmetric FIR filter structure to a generic FPGA architecture and meet the challenges regarding routability, timing, and precision.

\subsection{Prior work}
Several convenient structures of FIR filters can be found in the literature, e.g.\ direct-form realization~\cite{proakis}, transposed direct-form~\cite{meyerbaese}, and symmetry exploiting direct-form structures for linear-phase systems~\cite{oppenheim}.
The literary work covers the basic mathematical operations and structures but bypasses the technical aspects concerning the prevalent DSP block architectures of contemporary FPGAs. Nevertheless, the major manufacturers of FPGAs support their platforms with practical explanations~\cite{alterafir,xcell3,xcell4}.

Besides the hardware-mapped structures, an improved precision of a symmetric FIR filter without increasing the coefficient bit-widths was proposed by Shen~\cite{shen}. The presented parallel method implementation utilizes an accumulator in combination with variable shift operations. Although FPGAs have "flexible multidata bus routing capabilities" \cite{shen}, the suggested shift of the accumulator input contradicts the architecture of state-of-the-art embedded DSP blocks (e.g.\ \cite{altera_wp,dsp48e}). As Shen omits a benchmark with an FPGA implementation, the synthesis results of the optimized structure on a Xilinx FPGA were presented by Yuan~\cite{yuan}. These results show a further consumption of slice logic in addition to the DSP blocks, but neglect the specific architecture of the FPGA and do not properly include the DSP blocks in their optimization.

\subsection{Aim of this work}
High-order digital filters require a cascade of sub-filters for an efficient realization \protect\cite{mehrnia2} or, alternatively, an efficiently mapped hardware design. With regard to recent hardware architectures, a FIR filter whose length exceeds the number of cascaded DSP blocks (DSP chains) in an FPGA, can be referred to as high-order filter.
This brief presents the methodology and implementation of a systolic FIR filter in an FPGA, which ideally matches the prevalent embedded DSP block architecture. Consequently, the direct implementation of large structures avoids the cascading into small filters. The challenges of such a design are discussed, exemplary solved, and compared to the key performance indicators of an FPGA implementation: logic utilization and clock frequency.
Moreover, in reference to the "bit compression" introduced by Shen~\cite{shen}, we adapt their parallel method and implement an improved precision of the frequency response without additional logic utilization.

\section{Optimized FIR filter structure for FPGAs} 
\begin{figure}[ht]
\centering
\includegraphics[width=0.3\textwidth]{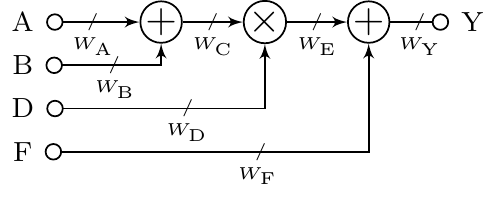}
\caption{The prevalent architecture for a DSP operation in an FPGA with pre-adder ($C=A+B$), multiplier ($E={C}\times{D}$) and post-adder ($Y=E+F$). The bit-widths $W$ of the inputs and outputs depend on the FPGA architecture.}
\label{fig_dspblock}
\end{figure}

Within the scope of high-order parallel FIR filters, it is suitable to design the filter with a linear phase. Thus, the coefficients $h[k]$ of the FIR system of order $M$ would satisfy the symmetry condition $h[M-k] = h[k]$, where $k = 0,1,...,M$ and the number $N$ of coefficients and taps for the FIR system is $N = M + 1$ \cite{oppenheim}. Consequently, the generalized equation for the output $y$ of a symmetric FIR filter with an even number $N$ of taps can be expressed as \cite{oppenheim}:
\begin{equation}
y[n] = \sum_{k=0}^{\frac{M-1}{2}} h[k] \left(x[n-k] + x[n-M+k]\right)\label{eqn_fir_sym} \, .
\end{equation}
In the case of symmetric FIR filters, the number of coefficient multipliers is essentially halved \cite{oppenheim}. Other types of symmetry, e.g.\ point symmetry or an odd number $N$ of taps, are developed by an analogy with the ongoing methodology. However, the taps of the input samples $x[n]$ are folded around half of the filter length by ${N}/{2}$ pre-adders, before the sums are multiplied by $h[k]$. Finally, the products are convoluted by ${N}/{2}-1$ post-adders. Hence, a basic arithmetic logic unit for a DSP operation incorporates a pre-adder, a multiplier and a post-adder as shown in Fig.~\protect\ref{fig_dspblock}.

\begin{figure}[ht]
\centering
\includegraphics[width=0.49\textwidth]{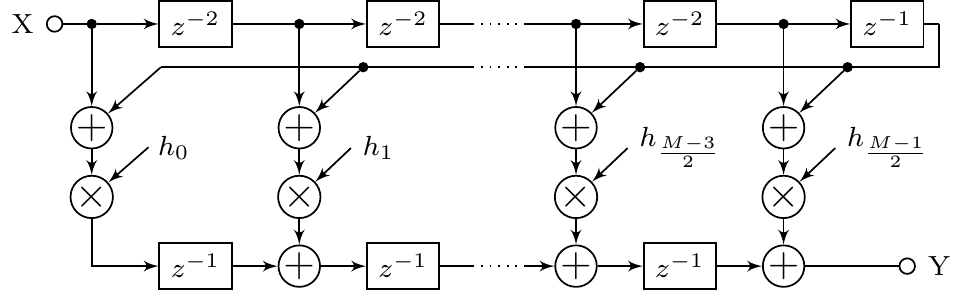}
\caption{A linear phase FIR filter of order $M$ (odd number). The filter function is folded around half of the filter length $N$ (even number) according to Eq.~\protect\eqref{eqn_fir_z}, and the adder tree for the running sum is replaced by the systolic structure derived from Eq.~\protect\eqref{eqn_rec_pipe}.}
\label{fig_systolic_maxdelay}
\end{figure}

Contemporary FPGA architectures embed multiply-accumulate blocks for digital filters, either by integrated hardware primitives or by synthesized logic blocks. The mapping of Eq.~\protect\eqref{eqn_fir_sym} to parallel DSP blocks according to the direct-form implementation is convenient, but results in a large adder tree for the running sum, which slows down the maximum clock frequency of the system. If the design targets a high throughput at the highest clock frequency, a systolic structure with a pipeline will maximize the performance of the FIR filter, as long as latency is negligible. The structure is described in $z$-domain by using the time-shifting property of the $z$-transformation $x[n-k] \stackrel{z}{\leftrightarrow} z^{-k}X(z)$~\protect\cite{oppenheim} and Eq.~\protect\eqref{eqn_fir_sym} resulting in:
\begin{equation}
\label{eqn_fir_z}
Y = \sum_{k=0}^{\frac{M-1}{2}} h_k \left(z^{-k} + z^{-M+k}\right) X\, ,
\end{equation}
where $h_k = h[k]$. Furthermore, a pipeline is embedded into the systolic structure by adding registers to the output of each running sum element. In terms of the $z$-transformation, the pipeline register is a unit delay $z^{-1}$. Therefore, the pipelining of the FIR filter can be expressed as:
\begin{align}
Y z^{-1}=& z^{-1} \sum_{k=0}^{\frac{M-3}{2}} h_k \left(z^{-k} + z^{-M+k} \right)X + \nonumber\\
         & z^{-1} h_{\frac{M-1}{2}} \left( z^{\frac{-M+1}{2}} + z^{\frac{-M-1}{2}} \right)X\label{eqn_pipeone}\\
Y z^{-2}=& z^{-1} \Bigg[z^{-1} \sum_{k=0}^{\frac{M-5}{2}} h_k \left(z^{-k} + z^{-M+k} \right)X  +\nonumber\\
         & \phantom{z^{-1} \Bigg[}z^{-1} h_{\frac{M-3}{2}} \left( z^{\frac{-M+3}{2}} + z^{\frac{-M-3}{2}} \right)X\Bigg] + \nonumber\\
         & z^{-2} h_{\frac{M-1}{2}} \left( z^{\frac{-M+1}{2}} + z^{\frac{-M-1}{2}} \right)X\label{eqn_pipetwo} \, .
\end{align}
Eq.~\protect\eqref{eqn_pipeone} illustrates the decomposition of the last term of the running sum with a pipeline register. In addition, the second pipeline stage is formed by Eq.~\protect\eqref{eqn_pipetwo}. Finally, the complete decomposition of the adder tree with $\frac{M-1}{2}$ unit delays reveals the FIR filter function in terms of iteration of the basic systolic element
\begin{align}
Y_k &= z^{-1} Y_{k-1} + z^{-k} h_k \left( z^{-k} + z^{-M+k}\right)X\nonumber\\
    &= z^{-1} Y_{k-1} + h_k \left( z^{-2k} + z^{-M}\right)X \label{eqn_rec_pipe}\\ 
Y_0 &= h_0 \left( 1 + z^{-M} \right)X\label{eqn_rec_pipe_init}\, ,                                          
\end{align}
where the output $Y$ from Eq.~\protect\eqref{eqn_fir_z} is equal to the output of the last pipeline stage at position $k = \frac{M-1}{2}$ from Eq.~\protect\eqref{eqn_rec_pipe}. The block diagram of the symmetric systolic FIR filter is shown in Fig.~\protect\ref{fig_systolic_maxdelay}. A detailed mapping of this structure to the Xilinx specific architecture is shown in \protect\cite{xilinx_dsp}.

\begin{figure}[t]
\centering
\includegraphics[width=0.49\textwidth]{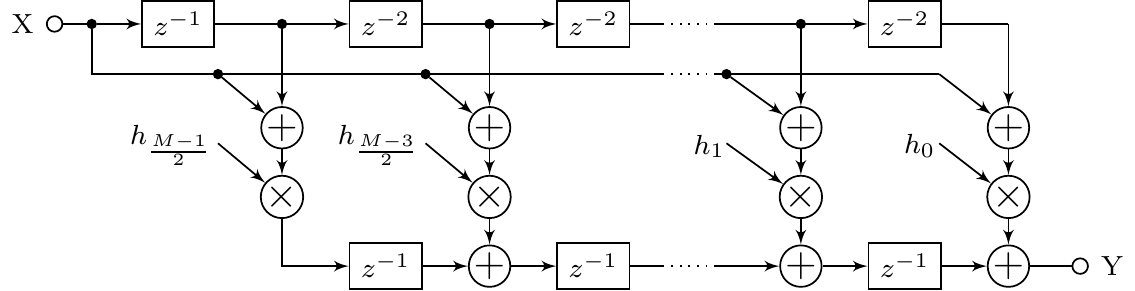}
\caption{A block diagram of the FIR filter structure from Eq.~\protect\eqref{eqn_rec_pipe_fast} with reduced initial pipeline delay.}
\label{fig_systolic_mindelay}
\end{figure}

It is clear, that the pipeline registers at the outputs of the multiply-accumulate operations cause a delay of $k$ clock cycles, but the derived iteration from Eq.~\protect\eqref{eqn_rec_pipe} also inserts an initial pipeline delay, as the first folded sum from Eq. \protect\eqref{eqn_rec_pipe_init} is only valid after passing the tapped delay line with $M$ stages. A modification of the fundamental filter function from Eq.~\protect\eqref{eqn_fir_z} brings it to
\begin{equation}
Y = \sum_{k=0}^{\frac{M-1}{2}} h_{\frac{M-1}{2}-k} \left(z^{k} + z^{-1-k}\right) X \, ,
\label{eqn_fir_fast}
\end{equation}
and yields an alternative iterated function
\begin{align}
Y_k &= z^{-1} Y_{k-1} + h_{\frac{M-1}{2}-k} \left( 1 + z^{-1-2k}\right)X \label{eqn_rec_pipe_fast}\\ 
Y_0 &= h_{\frac{M-1}{2}-k} \left( 1 + z^{-1} \right)X\label{eqn_rec_pipe__fast_init}\, ,  
\end{align}
where the initial delay is reduced to the unit delay. The corresponding block diagram is shown in Fig.~\protect\ref{fig_systolic_mindelay}. The overall latency is determined by the number of DSP blocks, which in this case is $N/2$.

The featured symmetric systolic structure is generic so as to match the DSP block architecture of state-of-the-art FPGAs. Although the vendors put forward the systolic FIR filter, the synthesis of the structure faces two major challenges, discussed in the following subsections~\ref{sec_routability} and \ref{sec_timing}. The achieved results are highlighted in section~\protect\ref{sec_results_structure}
\subsection{Routability}
\label{sec_routability}
Firstly, as the DSP blocks are embedded at dedicated locations in multiple chain-like structures, the interconnection capabilities are limited. In correlation with the filter order, the routability experiences less flexibility. Thus, successful mapping and routing of relatively large filters, compared to the number of DSP blocks, mainly depends on the ability of the development tools to exploit the DSP chain architecture. Even though a generic VHDL design permits an unrestricted synthesis, several failures were observed during the implementation process. Actually, physically separated DSP chains limit a realization of high-order FIR filters beyond the length of a DSP chain. An efficient concatenation of DSP chains is impossible for the evaluated tools without adaptation. Moreover, the mapping of the systolic structure to dedicated slices becomes more error prone, as the number of multiply-accumulate operations exceeds the total number of available DSP blocks. In this case, the tools cannot map the systolic FIR filter to the DSP blocks and distributed arithmetic built on the slice logic at the same time. In conclusion, the implementation of systolic FIR filters, which overlap multiple DSP chains, requires further efforts in the design. As a manual placement and routing is inadequate, the generic structure of the filter is adapted in such a manner so as to support the routability.

\begin{figure}[t]
\centering
\includegraphics[width=0.49\textwidth]{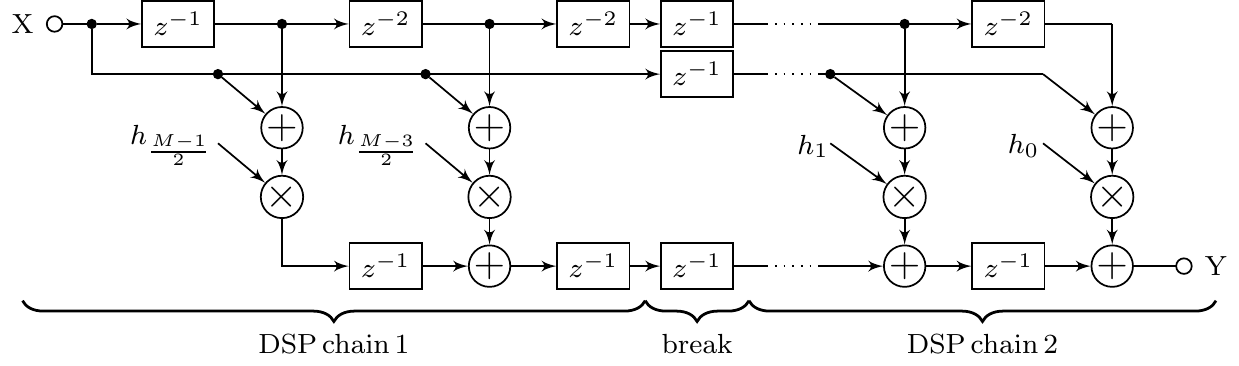}
\caption{The symmetric systolic FIR structure with additional registers for improved routability and timing. Registers between the systolic elements break the dedicated routes of DSP chains.}
\label{fig_systolic_firbreak}
\end{figure}

The routability is rapidly improved by further pipeline registers between the cascaded running sum. Consequently, the iterated function from Eq.~\protect\eqref{eqn_rec_pipe_fast} ultimately changes to
\begin{numcases}{Y_k=}
   z^{-1} Y_{k-1} + h_{\frac{M-1}{2}-k} \left( z^{-b_k} + z^{-1-2k-b_k}\right)X \label{eqn_rec_pipe_break}
   \\
   z^{-1} Y_{k-1} \label{eqn_pipe_break} \, ,
\end{numcases}
where Eq.~\protect\eqref{eqn_pipe_break} covers the case that a register is inserted at an arbitrary position $k$ satisfying the condition $\frac{M-1}{2}>k>1$. The elements $b_k$ of Eq.~\protect\eqref{eqn_rec_pipe_break} represent the number of injected registers before position $k$. At least one register between the DSP blocks breaks the systolic structure to match a two-column chain architecture. Dedicated routes of DSP chains are thus replaced through slice logic utilizing a simple register (Fig.~\protect\ref{fig_systolic_firbreak}). Moreover, this method also facilitates the routing of the systolic FIR structure within DSP blocks in combination with distributed arithmetic. After all of the above, the number of DSP blocks or the length of DSP chains no longer restricts the order of realizable digital filters.

\subsection{Timing}
\label{sec_timing}
Secondly, within the scale of an FPGA, DSP chains are distantly located. Each type of device comes along with its specific physical dimensions and arrangement of configurable logic. However, the minimum clock period for the circuit primarily depends on the longest path between the logic elements. The interconnections of DSP chains can therefore be a bottleneck in terms of timing and thus limit the maximum achievable clock frequency. Even if the tools are capable of mapping systolic FIR filters spanning multiple DSP chains, the path lengths cannot be automatically reduced. In this case, a partial break with one register or more at distinguished positions in accordance with the DSP chain lengths reduces the overall path lengths (Fig.~\protect\ref{fig_systolic_firbreak}). The maximum clock frequency of the FPGA design is consequently the maximum sample rate of the filter. That limit is verified by timing constraints with the design tools.

\subsection{Results}
\label{sec_results_structure}
For the proof of concept, we chose an FPGA from Xilinx (Artix 7, XC7A35T-3CSG324)  and from Altera (Cyclone 5, 5CEFA5F23C6). Their DSP blocks \cite{altera_wp,dsp48e} are able to perform the basic operations from Fig.~\protect\ref{fig_dspblock}. The bit-widths of the generic VHDL design are adjusted to be $(W_A,W_B,W_C,W_D,W_E,W_F) = (15,15,16,18,34,36)$, where the 15-bit wide input samples ($W_A, W_B$) are adapted to our application, and, furthermore, the bit-widths of the multiplier ($W_C, W_D$) and post-adder ($W_E, W_F$) are chosen to avoid overflows and to fit into both FPGA architectures. To illustrate an example, a low-pass filter with a cutoff frequency $f_{\mathrm{pass}}$ to be one-tenth of the sampling frequency $f_\mathrm{s}$ and a stopband frequency $f_{\mathrm{stop}}$ to be one-eighth of $f_\mathrm{s}$ was selected. Furthermore, the stopband attenuation $A_\mathrm{stop}$ is chosen to be $102\,\mathrm{dB}$, which corresponds to the quantization noise floor of $18\,\mathrm{bit}$ signed coefficients. Estimating the number $N_{\mathrm{FIR}}$ of taps with \cite{lyons}
\begin{equation}
N_{\mathrm{FIR}} \approx \frac{A_\mathrm{stop}}{22 \left( f_{\mathrm{stop}} - f_{\mathrm{pass}} \right)} \, ,
\end{equation}
$186\,\text{taps}$ are required to achieve that frequency response. However, we truncated $N_{\mathrm{FIR}}$ to $180$, because the symmetry exploiting structure ultimately fits the total $90$ DSP blocks of the Xilinx device. The Altera device includes $150$ DSP blocks. 
Finally, the adaptation of the straightforward structure with dedicated register stages at distinct positions in the data path was evaluated and compared to the state-of-the-art FIR compiler tools \cite{xilinxfircompiler,alterafircompiler} in Tab.~\protect\ref{tab_result}.
\begin{table}[ht]
\renewcommand{\arraystretch}{1.1}
\centering
\caption{Exemplary logic utilization of 90-tap FIR filters}
\label{tab_result}
\begin{tabular}{p{2.2cm}p{2.6cm}p{2.6cm}}
\hline
{90 taps}                        &     {Vivado 2016.2}                                                             & {Quartus Prime 16.0} \\
{systolic structure}             &     {XC7A35T-3CSG324}                                                           & {5CEFA5F23C6} \\
\hline
\hline                                                              
{straightforward}                 &     {LUT$\footnotesize{{}^{1}}$: 5 / 20,800 \newline Reg$\footnotesize{{}^{3}}$.: 4 / 41,600 \newline DSP$\footnotesize{{}^{4}}$: 90 / 90}       & {ALM$\footnotesize{{}^{2}}$: 1,345 / 29,080 \newline Reg.: 2,537 / 58,160 \newline DSP: 76 / 150} \\
\hline                            
{partial break ($z^{-1}$)}        &     {LUT: 3 / 20,800 \newline Reg.: 18 / 41,600 \newline DSP: 90 / 90}          & {ALM: 1,292 / 29,080 \newline Reg.: 2,434 / 58,160 \newline DSP: 76 / 150} \\
\hline                            
{full break ($z^{-1}$)}           &     {LUT: 60 / 20,800 \newline Reg.: 152 / 41,600 \newline DSP: 90 / 90}        & {ALM: 188 / 29,080 \newline Reg.: 441 / 58,160  \newline DSP: 88 / 150} \\
\hline                            
{full break ($z^{-2}$)}           &     {LUT: 45 / 20,800 \newline Reg.: 3429 / 41,600 \newline DSP: 90 / 90}       & {ALM: 1,009 / 29,080  \newline Reg.: 3736 / 58,160  \newline DSP: 88 / 150} \\
\hline
{Xilinx\newline FIR Compiler 7.2\newline\cite{xilinxfircompiler}} &     {LUT: 3201 / 20,800 \newline Reg.: 4293 / 41,600 \newline DSP: 90 / 90}     & {\phantom{text}  not applicable} \\
\hline
{Altera\newline FIR Compiler 16.0\newline\cite{alterafircompiler}}&     {\phantom{text} not applicable}                                             & {ALM: 801 / 29,080  \newline Reg.: 2883 / 58,160  \newline DSP: 45 / 150} \\
\hline
\end{tabular}
\raggedright
\footnotesize{$\phantom{}^{\phantom{I}}$ $(1)$ look-up table, $(2)$ adaptive logic module (an ALM implements a LUT), $(3)$ register, $(4)$ DSP block with pre-adder, multiplier, and post-adder}
\end{table}

With identical configurations for the bit-widths, both tools were capable of implementing the systolic structure of the 90\,taps FIR filter. Apparently, a Xilinx implementation is capable of including the tapped delay line and the output registers into DSP blocks in various ways. Therefore, the tool maps the investigated structures efficiently to the hardware architecture, consuming very less ($<\,0.5\%$) additional logic. On the contrary, the DSP architecture of Altera is not that versatile, as only one variant of the FIR filter structure results in competitive results with less than $1\,\%$ additional logic. As a result, the structure with one additional register stage at the output of each systolic element performs best in terms of resource utilization. 

As discussed, the timing performance is improved by further registers breaking the dedicated routes. The obtained values from the tools are shown in Tab.~\protect\ref{tab_result_timing}.
\begin{table}[ht]
\renewcommand{\arraystretch}{1.2}
\centering
\caption{Maximum achievable clock frequencies of 90-tap FIR filters}
\label{tab_result_timing}
\begin{tabular}{lcc}
\hline
{90 taps}                        & {Vivado 2016.2}        & {Quartus Prime 16.0} \\
{systolic structure}             & {XC7A35T-3CSG324}      & {5CEFA5F23C6}        \\
\hline
\hline
{straightforward}                & {$238.10\,\mathrm{MHz}$}  & {$145.65\,\mathrm{MHz}$}\\
{partial break ($z^{-1}$)}       & {$303.03\,\mathrm{MHz}$}  & {$148.82\,\mathrm{MHz}$}\\ 
{full break ($z^{-1}$)}          & {$476.19\,\mathrm{MHz}$}  & {$218.77\,\mathrm{MHz}$}\\
{full break ($z^{-2}$)}          & {$526.32\,\mathrm{MHz}$}  & {$231.75\,\mathrm{MHz}$}\\
{Xilinx FIR Compiler}            & {$434.78\,\mathrm{MHz}$}  & {not applicable}        \\ 
{Altera FIR Compiler}            & {not applicable}          & {$213.72\,\mathrm{MHz}$}\\ 
\hline
\end{tabular}
\end{table}

The best timing performance is achieved by the fully pipelined structure with two additional registers at each output. Further registers have no noticeable impact on timing. The values for the clock frequencies were obtained from the slow process corner of the timing reports. For the Xilinx implementation, the timing constraints were successively increased to reach the limit. Besides the improved timing in terms of maximum clock frequency, the overall latency of the filter is increased depending on the injected register stages.

In conclusion, at least one variant of our generic design utilizes less logic and operates at a higher clock frequency than the compiler-generated implementation. The efficiency is caused by the simplicity of the approach, whereas the automated design tools include excessive configuration options.

\section{Optimized fixed-point arithmetic}
For the implementation of high-order FIR filters, the coefficient quantization becomes significant, as the granularity of coefficients increases. Thus, an increased dynamic range for the quantization of coefficients preserves the accuracy of the digital filter. The precision of a signed fixed-point operation depends on the bit-width $b$ supported by the hardware. Hence, the transformation from a real coefficient $h$ to the corresponding integer value $I \in \mathbb{Z}$, used for a signed fixed-point calculation is, in general, described as:
\begin{equation}
I = \mathrm{round}{ \left( h 2^{b-1} \right)} \longleftrightarrow \frac{I}{2^{b-1}}  = h_I \, . 
\label{eqn_fixedpoint}
\end{equation}
An efficient method for an improved fixed-point precision was shown by Shen \cite{shen}. This method, referred to as "bit compression", performs a left shift operation on the integer value, until all redundant sign extension bits are removed. However, this equals a multiplication with $2^{Q}$, and we calculate $Q$ as follows:
\begin{equation}
Q = \left\lfloor \mathrm{log}_{2} \left( \frac{2^{b-1}}{\left| h \right|} \right) - (b - 1) \right\rfloor \, .
\label{eqn_bitcompression}
\end{equation}
The removal of sign extension bits increases the dynamic range of the fixed-point representation of the coefficients, but requires that all products are normalized to a common base before they are added by the accumulator. Therefore, Shen's parallel method implementation performs a normalization at the input of the running sum accumulator. Indeed, the operation is incompatible to the DSP block architecture and is therefore synthesized on distributed logic. Thus, the proposed FIR filter structure does not exploit the entire performance of a hardware mapped systolic FIR filter.

\begin{figure}[ht]
\centering
\includegraphics[width=0.49\textwidth]{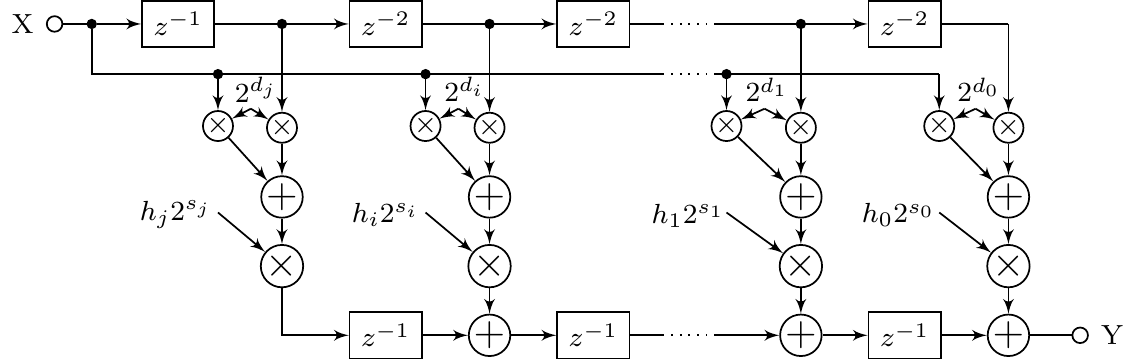}
\caption{A block diagram of the systolic FIR filter with enhanced dynamic range for fixed-point coefficients by including left shift operations.}
\label{fig_firshift}
\end{figure}

In general, the normalization of a partial term $S$ from Eq.~\protect\eqref{eqn_fir_fast} to a common base is realized in fixed-point representation by:
\begin{equation}
\label{eqn_normalization}
2^{-d_k} S = \mathrm{round}{ \left( h_k 2^{b - 1 + Q_k} \right) } (z^{-k} + z^{-1-k}) X \,
\end{equation}
where $Q_k$ is calculated by Eq.~\protect\eqref{eqn_bitcompression} and $d_k$ is a normalization factor, which must be individually calculated for each coefficient. Moreover, Eq.~\protect\ref{eqn_normalization} reveals that the multiplication with $2^{d_k}$ (left~shift operation) performs normalization and can be applied to the delayed input samples (Fig.~\protect\ref{fig_firshift}).
For our approach, the $d_k$ left~shift operation for normalization is restricted by the bit-widths of the DSP block architecture, and therefore $Q_k$ must be limited to an upper value. The bounded value $Q_k^{'}$ is determined by the largest bit shift applicable to the input samples but does not exceed the bit-width $W_C$ of the pre-adder. That method benefits from a generously designed pre-adder with regard to the bit-width. With a look at the contemporary DSP architecture from Xilinx \protect\cite{dsp48e}, the pre-adder operates at $W_C = 25\,\mathrm{bit}$ and the multiplier supports $25\,\mathrm{bit} \times 18\,\mathrm{bit}$ operations. Thus, for 16-bit input samples and a 25-bit wide pre-adder, the limit of $Q_k'$ is 9-bit.

\subsection{Results}

\begin{figure}[ht]
\centering
\includegraphics[width=0.49\textwidth]{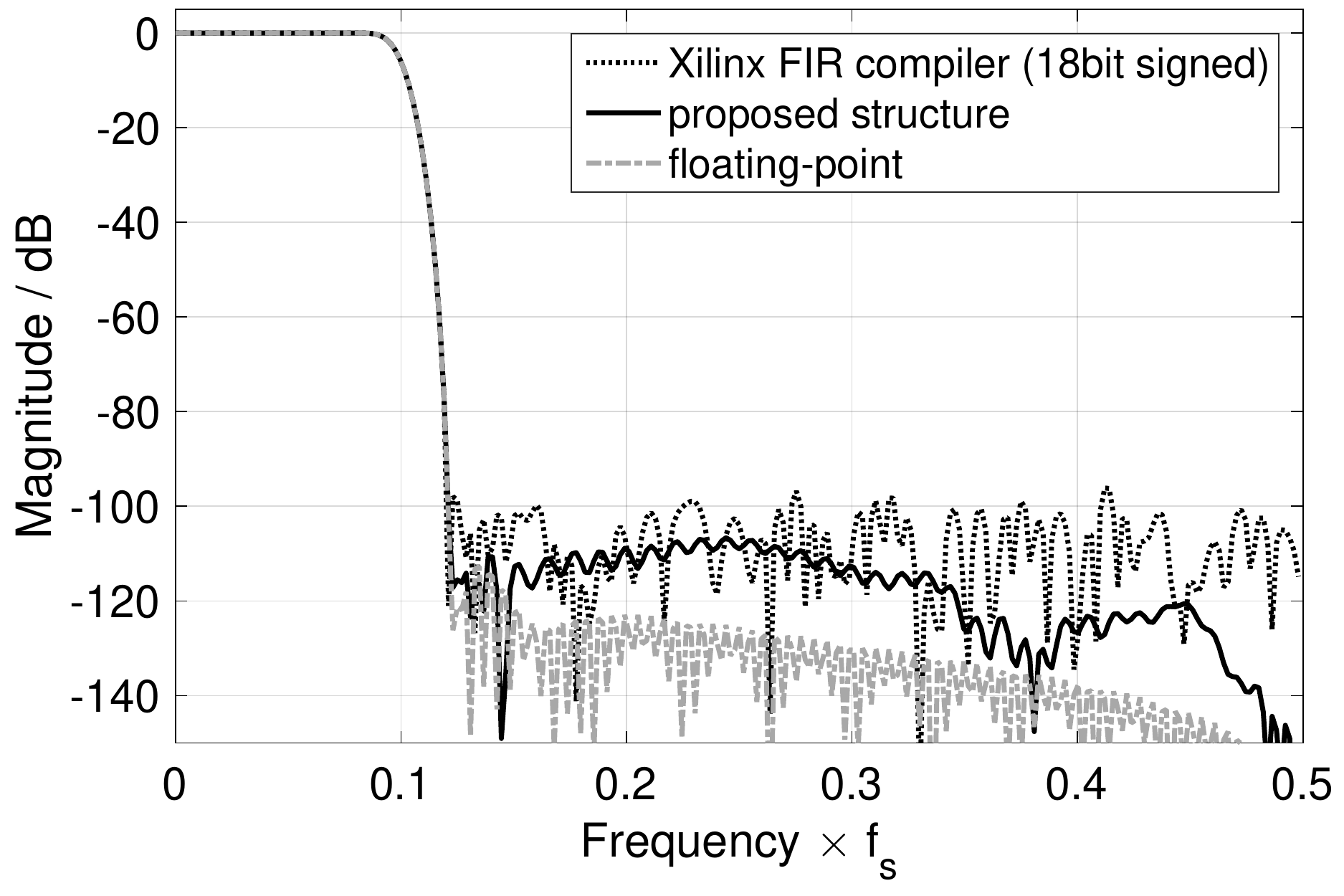}
\caption{The frequency responses of a symmetric systolic FIR filter with 90~taps depended on the numeric representation of the coefficients.}
\label{fig_frequency_response}
\end{figure}

For the evaluation of the proposed structure from Fig.~\protect\ref{fig_firshift}, we designed a low-pass filter of order 179 with the window method~\protect\cite{lyons} based on Nuttall's window \protect\cite{nuttall}. Furthermore, the result of the FIR filter generated by the Xilinx FIR Compiler is compared to that of the proposed structure with additional bit shift operations and a floating point calculation. The frequency responses, which are derived by a Fourier transform of the simulated impulse responses of the filters, are shown in Fig.~\protect\ref{fig_frequency_response}.

A comparison of frequency responses of the compiled FIR filter structures from Xilinx and Altera with $18\,\mathrm{bit}$ signed coefficients reveals no significant differences. However, our proposed structure with shift operations for coefficient normalization results in an improved stopband attenuation, even though it also exploits $18\,\mathrm{bit}$ signed coefficient multiplier. On a Xilinx FPGA, that structure utilizes exactly the same logic resources in comparison to the equivalent variant without left shift operations. The critical path delays remain constant without reducing the maximum clock frequency.

\section{Conclusion}
In this brief, a systolic structure for a symmetric FIR filter was proposed, where the systolic elements ideally match the prevalent DSP block architecture of FPGAs. Thus, the derived iterative mathematical functions for the systolic structure support the mapping of a digital filter to various architectures with different constraints. Moreover, a generic FIR filter design was synthesized by the tools from Altera and Xilinx to evaluate the efficiency of the structure in terms of logic utilization and maximum clock frequency. The results confirmed that the proposed structure with additional register stages improves routability and timing of high-order FIR filters and is superior to state-of-the-art FIR compiler tools. Furthermore, to yield an increased dynamic range for coefficients quantization, we enhanced the structure by shift operations, thus improving the precision of fixed-point arithmetic. Finally, the exploitation of the entire DSP blocks enables an efficient realization of high-order FIR filters with fixed-point arithmetic in FPGAs, while utilizing less than $1\,\%$ additional slice logic and running at clock frequencies above $200\,\mathrm{MHz}$.


\begin{thebibliography}{1}

\bibitem{meher}
P. K. Meher, S. Chandrasekaran, A. Amira, {"FPGA Realization of FIR Filters by Efficient and Flexible Systolization Using Distributed Arithmetic"}, \emph{IEEE Transactions on Signal Processing}, vol.~56, no.~7, pp.~3009-3017, July 2008.

\bibitem{park}
S. Y. Park, P. K. Meher, {"Efficient FPGA and ASIC Realizations of a DA-Based Reconfigurable FIR Digital Filter"}, \emph{IEEE Transactions on Circuits and Systems II: Express Briefs}, vol.~61, no.~7, pp.~511-515, July~2014.


\bibitem{mehrnia2}
A. Mehrnia, A. N. Willson, {"FIR Filter Design Using Optimal Factoring: A Walkthrough and Summary of Benefits"}, \emph{IEEE Circuits and Systems Magazine}, vol.~16, no.~1, pp.~8-21, 2016.

\bibitem{xilinx_dsp}
Xilinx, Inc., \emph{DSP: Designing for Optimal Results}, Xcell Publications, Edition 1.0, April 2005, p.~39, p.~80 .

\bibitem{altera_wp}
Altera Corporation, \emph{Enabling High-Performance DSP Applications with Arria V or Cyclone V Variable-Precision DSP Blocks}, Altera White Paper WP-01159-1.0, 2011, pp.~4-6.

\bibitem{dsp48e}
Xilinx, Inc., \emph{7 Series DSP48E1 Slice}, User Guide, UG479, v1.8, November 2014, p.~14.

\bibitem{proakis}
{J. G. Proakis, D. K. Manolakis}, \emph{Digital Signal Processing}, Fourth Edition, Pearson New International Edition, Pearson Education Limited, 2013, p.~580.

\bibitem{meyerbaese}
{U. Meyer-Baese}, \emph{Digital Signal Processing with Field Programmable Gate Arrays}, Springer Publishing Company, Inc., 3rd Edition, 2007, pp.~166-167.

\bibitem{oppenheim}
{A. V. Oppenheim and R. W. Schafer}, \emph{Discrete-Time Signal Processing}, Third Editon, Prentice Hall, 2010, p.~154, p.~431.

\bibitem{alterafir}
Altera Corporation, \emph{Implementing FIR Filters and FFTs with 28-nm Variable-Precision DSP Architecture}, Altera White~Paper WP-01140-1.0, 2010.



\bibitem{xcell3}
A. P. Taylor, {"Ins and Outs of Digital Filter Design and Implementation"}, \emph{Xcell Journal}, issue~78, pp.~36-41, 2012.

\bibitem{xcell4}
R. Zatrepalek, {"Using FPGAs to Solve Tough DSP Design Challenges"}, \emph{Xcell Journal}, issue~78, pp.~42-47, 2012.


\bibitem{shen}
Z. Shen, {"Improving FIR Filter Coefficient Precision [DSP Tips \& Tricks]"}, \emph{IEEE Signal Processing Magazine}, vol.~27, no.~4, pp.~120-124, July~2010.

\bibitem{yuan}
{J. Yuan,  Q. Y. Feng, D. Wang}, {"Design of High-Precision FIR Filter Based on Verilog HDL"}, \emph{Proc. MSIT}, Feb.~2012, pp.~5198-5202.


\bibitem{lyons}
{R. G. Lyons}, \emph{Understanding Digital Signal Processing}, Prentice Hall, 3rd Edition, 2010, pp.~186-235.

\bibitem{xilinxfircompiler}
{Xilinx, Inc.}, "FIR Compiler v7.2", PG149 November 18, 2015.

\bibitem{alterafircompiler}
{Altera Corporation}, "FIR II IP Core", UG-01072 October 01, 2015.

\bibitem{nuttall}
{A. Nuttall}, "Some windows with very good sidelobe behavior", \emph{IEEE Transactions on Acoustics, Speech, and Signal Processing}, vol.~29, no.~1, pp.~84-91, Feb~1981.

\end{thebibliography}
\end{document}